# Investigation of H Sorption and Corrosion Properties of $Sm_2Mn_xNi_{7-x}$ ($0 \leq x < 0.5$) Intermetallic Compounds Forming Reversible Hydrides

**Nicolas Madern, Véronique Charbonnier †, Judith Monnier, Junxian Zhang, Valérie Paul-Boncour and Michel Latroche ***

Paris Est Creteil University, CNRS, ICMPE, UMR7182, F-94320 Thiais, France; nicolas.madern@cea.fr (N.M.); v.charbonnier@aist.go.jp (V.C.); monnier@icmpe.cnrs.fr (J.M.); junxian@icmpe.cnrs.fr (J.Z.); paulbon@icmpe.cnrs.fr (V.P.-B.)

* Correspondence:latroche@icmpe.cnrs.fr

† Current affiliation: Energy Process Research Institute, National Institute of Advanced Industrial Science and Technology (AIST), Tsukuba West, 16-1 Onogawa, Tsukuba, Ibaraki 305-8569, Japan.

**Abstract:** Intermetallic compounds are key materials for energy transition as they form reversible hydrides that can be used for solid state hydrogen storage or as anodes in batteries. $AB_y$ compounds ($A$ = Rare Earth (RE); $B$ = transition metal; $2 < y < 5$) are good candidates to fulfill the required properties for practical applications. They can be described as stacking of [$AB_5$] and 2[$AB_2$] sub-units along the $c$ crystallographic axis. The latter sub-unit brings a larger capacity, while the former one provides a better cycling stability. However, $AB_y$ binaries do not show good enough properties for applications. Upon hydrogenation, they exhibit multiplateau behavior and poor reversibility, attributed to H-induced amorphization. These drawbacks can be overcome by chemical substitutions on the $A$ and/or the $B$ sites leading to stabilized reversible hydrides. The present work focuses on the pseudo-binary $Sm_2Mn_xNi_{7-x}$ system ($0 \leq x < 0.5$). The structural, thermodynamic and corrosion properties are analyzed and interpreted by means of X-ray diffraction, chemical analysis, scanning electron microscopy, thermogravimetric analysis and magnetic measurements. Unexpected cell parameter variations are reported and interpreted regarding possible formation of structural defects and uneven Mn distribution within the Ni sublattice. Reversible capacity is improved for $x > 0.3$ leading to larger and flatter isotherm curves, allowing for reversible capacity >1.4 wt %. Regarding corrosion, the binary compound corrodes in alkaline medium to form rare earth hydroxide and nanoporous nickel. As for the Mn-substituted compounds, a new corrosion product is formed in addition to those above mentioned, as manganese initiates a sacrificial anode mechanism taking place at the early corrosion stage.

## 1. Introduction

The energy demand for mobile and stationary applications is continuously growing in line with industrial and social development. To better manage energy sources and preserve Earth from climate change, numerous efforts are being made worldwide to develop renewable energies. Though a good alternative to fossil fuels, these renewable energy sources are intermittent and unevenly distributed. This leads to the development of new facilities offering large energy storage capacities using safe and cheap solutions. Metallic hydrides ($MH$) are smart materials able to store energy in a solid form either by dihydrogen gas sorption in pressurized tanks or by electrochemical conversion in alkaline or Li-ion batteries [1–10]. Among the large families of $MH$, few of them work close to ambient conditions (i.e., atmospheric pressure and room temperature (RT)). Basically, they can be described as binary $AB_y$ compounds ($1 \leq y \leq 5$), $A$ being an element forming very stable hydride at RT, whereas, at RT, $B$ can only form hydride at high pressure.

$AB_5$-type materials ($A$ = Rare Earth and $B$ = transition metal) have been successfully developed as anodes in Ni-*M*H batteries [8,9,11] or recently, as solid storage materials in hydrogen tanks for a resort energy supply, combined with an electrolyzer and a fuel cell [12]. However, these $AB_5$-type materials have limited weight energy density, as their high molar mass remain a penalty, despite their very high volumetric energy density. To overcome this drawback, beside $AB_5$ other binary systems ($A_5B_{19}$, $A_2B_7$, $AB_3$…) are investigated for the improvement of storage capacity. These $AB_y$ alloys ($2 < y < 5$) are extensively studied regarding their enhanced storage capacity [13,14]. These compounds generally form two polymorphs, crystallizing either in $R\bar{3}m$ ($R$) or $P6_3/mmc$ ($H$) space groups. Their structures can be described [8,15] as the stacking along the ***c*** crystallographic axis of two different sub-units: [$AB_2$] and [$AB_5$]. These stacking structures are expected to take advantage of the properties of the $AB_2$ compounds which have a larger sorption capacity [16] though lower crystalline stability [17], whereas the $AB_5$ compounds have lower capacities, but improved cycling stability. Indeed, several studies attest for the better performance of these types of materials used as anodes in alkaline batteries [18,19]. However, $AB_y$ binary compounds often exhibit complex properties combining multiplateau isotherms curves, large hysteresis, and poor reversibility related to amorphization upon cycling. These drawbacks can be solved by proper chemical substitutions either on the $A$ and/or the $B$ site. As an example, Mg for $A$ substitution has been shown to be very efficient to mitigate the multiplateau issue, leading to a large single and flat plateau at ambient conditions [20–22].

In the present work, we demonstrate that very similar behaviors can be achieved by substituting Mn for Ni in the pseudo-binary $Sm_2Mn_xNi_{7-x}$ system ($0 \leq x \leq 0.49$), leading to materials presenting practical properties using noncritical elements (Mn, Ni, Sm). This paper first describes the structural and thermodynamical properties of the system regarding hydrogen sorption. In the second section, the resistance to corrosion, a key parameter for these materials used in wet alkaline media, is thoroughly investigated. These experiments allow for discussing the capability of these materials to be used for energy storage.

## 2. Experimental Part

All compounds were synthesized from high purity elements (Sm (Alfa Aesar 99.9%), Ni (Praxair 99.95%) and Mn (Alfa Aesar 99.99%)). The elements were first melted in an induction furnace under secondary vacuum at high temperature under argon ($P_{Ar} \approx 40$ kPa, Linde 99.995%). This step was repeated three times to ensure good homogeneity. Small excess of Sm (3.5 wt %) and Mn (3 wt %) were initially added to compensate the losses induced by evaporation during these three steps. After induction melting, the ingots were crushed into powder (<100 µm), pressed into 2 g-pellets, wrapped in tantalum foil and heat-treated under argon atmosphere in a sealed stainless-steel crucible for 3 days at 800 °C, except for $Sm_2Ni_7$ which was annealed at 950 °C. All compounds were finally water quenched to preserve their high temperature structure.

Structural properties were characterized by X-ray diffraction (XRD) using a Bruker D8 DAVINCI diffractometer with Cu-$K_\alpha$ radiation, in a $2\theta$-range from 10 to 110° with a step size of 0.02°. Experimental data were analyzed by the Rietveld method using either FullProf [23] or TOPAS V4.2 [24] programs. For the $A_2B_7$ phase, atomic positions were kept fixed to those reported for their respective prototypes ($H$ or $R$) [25]. Due to the low contrast between manganese and nickel, Mn atoms were evenly distributed within the Ni sites, assuming compositional ratio ($x/7$) for site occupancies. Samarium occupancies were refined to check for off-stoichiometry. Sm-occupancies were kept equal in each site of each polymorph. Finally, the possible occurrence of stacking faults commonly observed for these types of compounds [26] was not considered, though some bumpy background in the two-theta range 33–35° attests for their possible presence.

Chemical compositions were determined by Electron Probe Micro-Analysis (EPMA), using a CAMECA SX-100. Prior to analysis, the samples were polished using 3-µm diamond abrasive paste and ethanol solvent on a woven tape.

*P-c* isotherm curves were measured at 25 °C by the Sieverts' method using 100-µm crushed powders. After each *P-c* measurement, hydrogenated samples were desorbed under dynamic primary vacuum at 150 °C. XRD measurements were also performed on dehydrogenated samples.

Corrosion properties were studied on samples crushed into powder [36 µm; 100 µm] and sampled in several 200-mg batches. Each of them was introduced in a different sample holder, immersed in KOH solution (8.7 M) and introduced in an oven at 25 °C. After different corrosion times (24 h, 1 w, 4 w, 8 w and 18 w; h = hour, w = week), KOH solution was removed with a Pasteur pipette. The powder was rinsed several times with a low concentrated KOH solution ($10^{-2}$ M) having a pH high enough to avoid any dissolution of the corrosion products. The powder was then dried for 24 h under primary vacuum at 40 °C and finally characterized by XRD and magnetic measurements.

A Zeiss LEO Scanning Electron Microscope (SEM) equipped with a field emission gun (FEG) and an energy dispersive X-ray detector (EDX) was used to study the layer morphology and chemical element distribution. Prior to analysis, the samples were covered with a protective and conductive 3-nm layer of Pt/Pd.

Raman micro-spectroscopy was performed with a LaBRAM HR 800 (Jobin-Yvon-Horiba) Raman micro-spectrometer equipped with a charge coupled device (CCD) detector. A He:Ne laser (632.81 nm) was used as the excitation source. The spectra were measured in back-scattering geometry with a resolution of about 0.5 $cm^{-1}$.

Magnetic measurements were recorded with a Physical Properties Measurement System (PPMS) from Quantum Design. Small sample amounts (typically 20–50 mg) were used either as a massive sample (pristine alloy) or a corroded powder immobilized with a resin in a Teflon sample holder. The samples were placed in a gelatin capsule and fixed with glass wool. Diamagnetic contribution from the sample holder was first measured and subtracted from the sample magnetization. Isotherm magnetization curves were measured at 300 K in an applied field range from 0 to 9 T. Amount of free metallic nanoporous nickel (np-Ni) in the corroded samples was calculated by extrapolation of the magnetization curves to zero field assuming that the ferromagnetic contribution comes exclusively from nickel (neglecting the possible Mn alloying effect, due to the very small Mn percentage). The corresponding mass percentage of Ni was obtained by normalization with the saturation magnetization of nanoporous nickel isolated from a fully corroded sample ($M_s$ = 39.9 $Am^2.kg^{-1}$ at 300 K compared to 53 $Am^2.kg^{-1}$ for bulk Ni) [27].

Thermal gravimetric analysis (TGA from Setaram-Setsys Evolution) experiments were carried out with around 20 mg of corroded powders introduced in a platinum/rhodium crucible. After 12 to 24 h of outgassing under primary vacuum ($10^{-2}$ Pa), the samples were heated from room temperature to 1000 °C at 5 $°C.min^{-1}$ in a 20 $mL.min^{-1}$ argon flow (Linde 99.999%). This thermal profile was repeated a second time to be used as a calibration curve. For the corroded samples, the weight losses were attributed to dehydration of the *RE* hydroxide following Equations (1) and (2) [28]:

$$2\,RE(OH)_3 \rightarrow 2RE OOH + 2H_2O \quad (1)$$

$$2\,RE OOH \rightarrow RE_2O_3 + H_2O \quad (2)$$

Thus, TGA measurements allow for estimating the relative quantity of *RE* hydroxide $m_{OH}$. In addition, the formation of metallic nickel can be estimated (if one neglects the small contribution of Mn) from the overall corrosion reaction Equation (3) [27,29–31]:

$$RE_zMn_xNi_{7-x} + 3z\,H_2O \rightarrow z\,RE(OH)_3 + (7-x)\,Ni + \frac{3z}{2}\,H_2 \text{ (Mn neglected)}$$

$$m_{Ni}(wt\%) = \frac{m_{RE(OH)3}\,(wt\%)}{M_{RE(OH)3}} * \left(\frac{7-x}{z}\right) * M_{Ni} \quad (3)$$

## 3. Results

### *3.1. Structural and Elemental Characterization of the Pristine Compounds*

Eight different $Sm_2Mn_xNi_{7-x}$ samples were melted and heat-treated varying the Mn amount ($x$ = 0, 0.12, 0.16, 0.26, 0.29, 0.33, 0.36, 0.49). Each sample was characterized by XRD and by EPMA. The average compositions for the 2:7 phases (either *R* or *H*) are given in Table 1. Full repartition of the EPMA measurements are shown in the Mn-Ni-Sm ternary phase diagram in Supplementary Materials (Figure S1).

XRD diagrams are presented in Figure 1. For each sample, the main phases are of $A_2B_7$-types (*R* or *H*). In some cases, the presence of weak contribution of $AB_5$ ($x$ = 0.26, 0.29) and some $AB_3$ secondary phases ($x$ = 0.36 and 0.49) is noticed. Beside the intermetallics, a few oxide traces (less than 6 wt % of either cubic of monoclinic $Sm_2O_3$) are also observed.

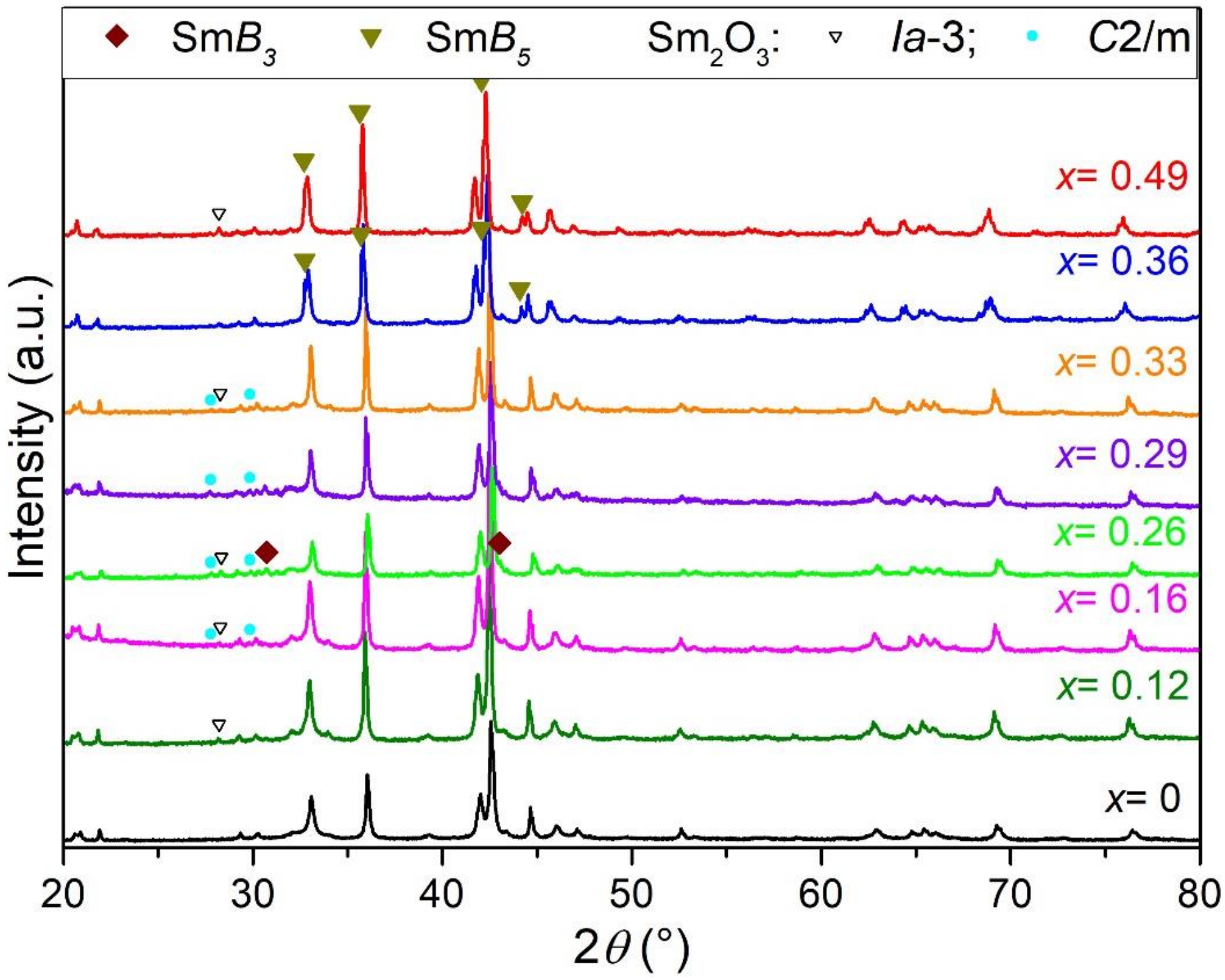

**Figure 1.** Evolution of the diffraction patterns of $Sm_2Mn_xNi_{7-x}$ as a function of x = 0, 0.12, 0.16, 0.26, 0.29, 0.33, 0.36, 0.49. One can notice the presence of $SmB_5$ and $SmB_3$ secondary phases (B = Mn, Ni) and traces of $Sm_2O_3$ oxides (either monoclinic C2/m or cubic Ia$\bar{3}$).

Rietveld refinements of the diffraction patterns are summarized in Table 1 (R factor, name of the phase, lattice parameters, R Bragg and phase abundance (%wt.)). All atomic positions were fixed to those published for $Sm_2Ni_7$ [24]. For all samples, the main 2:7 phase exists with the two polymorph structures (*H* and *R*). The relative ratio between these polymorphic phases ($\frac{H}{H+R}$) does not appear to be linked to the manganese content as observed in Table 1. It is more likely related to the reduced cell volume as shown in Figure 2 where a linear behavior is observed for all phases.

**Table 1.** Results of Electron Probe Micro-Analysis (EPMA) and XRD characterizations of $Sm_2Mn_xNi_{7-x}$ ($0 \le x \le 0.49$). All atomic positions were fixed to those published for $Sm_2Ni_7$ [25].

| $Sm_zMn_xNi_{7-x}$ (EPMA)*(z from XRD) | x | $R_{wp}$ | Phase | a | b | c | $R_B$ | Wt % (±1%) | $\frac{H}{H+R}$ (%) |
|---|---|---|---|---|---|---|---|---|---|
| $Sm_{2.01}Ni_7$ (z = 2.01(1)) | 0 | 20.1 | 2:7 (H) | 4.9811(3) | 4.9811(3) | 24.318(3) | 8.7 | 79.1 | 79.1 |
| | | | 2:7 (R) | 4.9792(4) | 4.9792(4) | 36.488(5) | 15.5 | 20.9 | |
| $Sm_{2.06}Mn_{0.12}Ni_{6.88}$ (z = 1.98(1)) | 0.12 | 2.9 | 2:7 (H) | 4.9876(3) | 4.9876(3) | 24.337(3) | 1.7 | 77.7 | 78.2 |
| | | | 2:7 (R) | 4.9857(2) | 4.9857(2) | 36.502(2) | 3.6 | 21.7 | |
| | | | $Sm_2O_3$ ($Ia\bar{3}$) | 10.939(4) | 10.939(4) | 10.939(4) | 2.2 | 0.6 | |
| $Sm_{1.94}Mn_{0.16}Ni_{6.84}$ (z = 1.97(1)) | 0.16 | 2.9 | 2:7 (H) | 4.9842(2) | 4.9842(2) | 24.312(2) | 2.5 | 73.2 | 76.2 |
| | | | 2:7 (R) | 4.9835(2) | 4.9835(2) | 36.469(3) | 3.9 | 22.9 | |
| | | | $Sm_2O_3$ (C2/m) | 14.13(2) | 3.641(3) | 8.893(8) | 1.8 | 3.1 | |
| | | | $Sm_2O_3$ ($Ia\bar{3}$) | 10.931(4) | 10.931(4) | 10.931(4) | 2.4 | 0.8 | |
| $Sm_{1.93}Mn_{0.26}Ni_{6.74}$ (z = 1.85(2)) | 0.26 | 20.4 | 2:7 (H) | 4.9794(3) | 4.9794(3) | 24.274(2) | 8.8 | 66.4 | 76.9 |
| | | | 2:7 (R) | 4.9776(7) | 4.9776(7) | 36.342(7) | 14.4 | 19.9 | |
| | | | $AB_5$ | 4.955(2) | 4.955(2) | 3.969(2) | 6.1 | 5.8 | |
| | | | $Sm_2O_3$ (C2/m) | 14.159(5) | 3.632(2) | 8.869(4) | 8.9 | 6.4 | |
| | | | $Sm_2O_3$ ($Ia\bar{3}$) | 10.932(2 | 10.932(2) | 10.932(2) | 12.1 | 1.5 | |
| $Sm_{1.92}Mn_{0.29}Ni_{6.71}$ (z = 1.91(2)) | 0.29 | 4.6 | 2:7(H) | 4.9816(3) | 4.9816(3) | 24.281(2) | 3.3 | 60.1 | 67.8 |
| | | | 2:7 (R) | 4.9787(6) | 4.9787(6) | 36.340(6) | 3.8 | 28.5 | |
| | | | $AB_5$ | 4.963(2) | 4.963(2) | 3.969(3) | 1.2 | 5.9 | |
| | | | $Sm_2O_3$ (C2/m) | 14.180(4) | 3.6226(8) | 8.855(2) | 1.5 | 5.5 | |
| $Sm_{1.91}Mn_{0.33}Ni_{6.67}$ (z = 1.96(1)) | 0.33 | 2.9 | 2:7 (H) | 4.9920(1) | 4.9920(1) | 24.329(1) | 3.2 | 78.3 | 81.0 |
| | | | 2:7 (R) | 4.9906(3) | 4.9906(3) | 36.483(3) | 4.8 | 18.4 | |
| | | | $Sm_2O_3$ (C2/m) | 14.15(2) | 3.640(3) | 8.888(9) | 1.0 | 2.8 | |
| | | | $Sm_2O_3$ ($Ia\bar{3}$) | 10.941(7) | 10.941(7) | 10.941(7) | 1.9 | 0.5 | |
| $Sm_{2.05}Mn_{0.36}Ni_{6.64}$ | 0.36 | 17.5 | 2:7 (H) | 4.9980(2) | 4.9980(2) | 24.360(2) | 9.8 | 62.0 | 81.6 |
| | | | 2:7 (R) | 4.9973(4) | 4.9973(4) | 36.542(4) | 14.6 | 14.0 | |
| | | | $AB_3$ | 5.0114(2) | 5.0114(2) | 24.549(2) | 9.4 | 24.0 | |
| $Sm_{1.97}Mn_{0.49}Ni_{6.51}$ (z = 1.99(1)) | 0.49 | 2.0 | 2:7 (H) | 5.0055(2) | 5.0055(2) | 24.389 (1) | 1.3 | 57.8 | 85.5 |
| | | | 2:7 (R) | 5.0051(6) | 5.0051(6) | 36.570(6) | 2.0 | 9.8 | |
| | | | $AB_3$ | 5.0144(2) | 5.0144(2) | 24.528(2) | 1.6 | 31.3 | |
| | | | $Sm_2O_3$ ($Ia\bar{3}$) | 10.938(2) | 10.938(2) | 10.938(2) | 1.8 | 1.1 | |

* x and z obtained from EPMA given directly in the formula $Sm_zMn_xNi_{7-x}$ (±0.02); z value from XRD below in brackets.

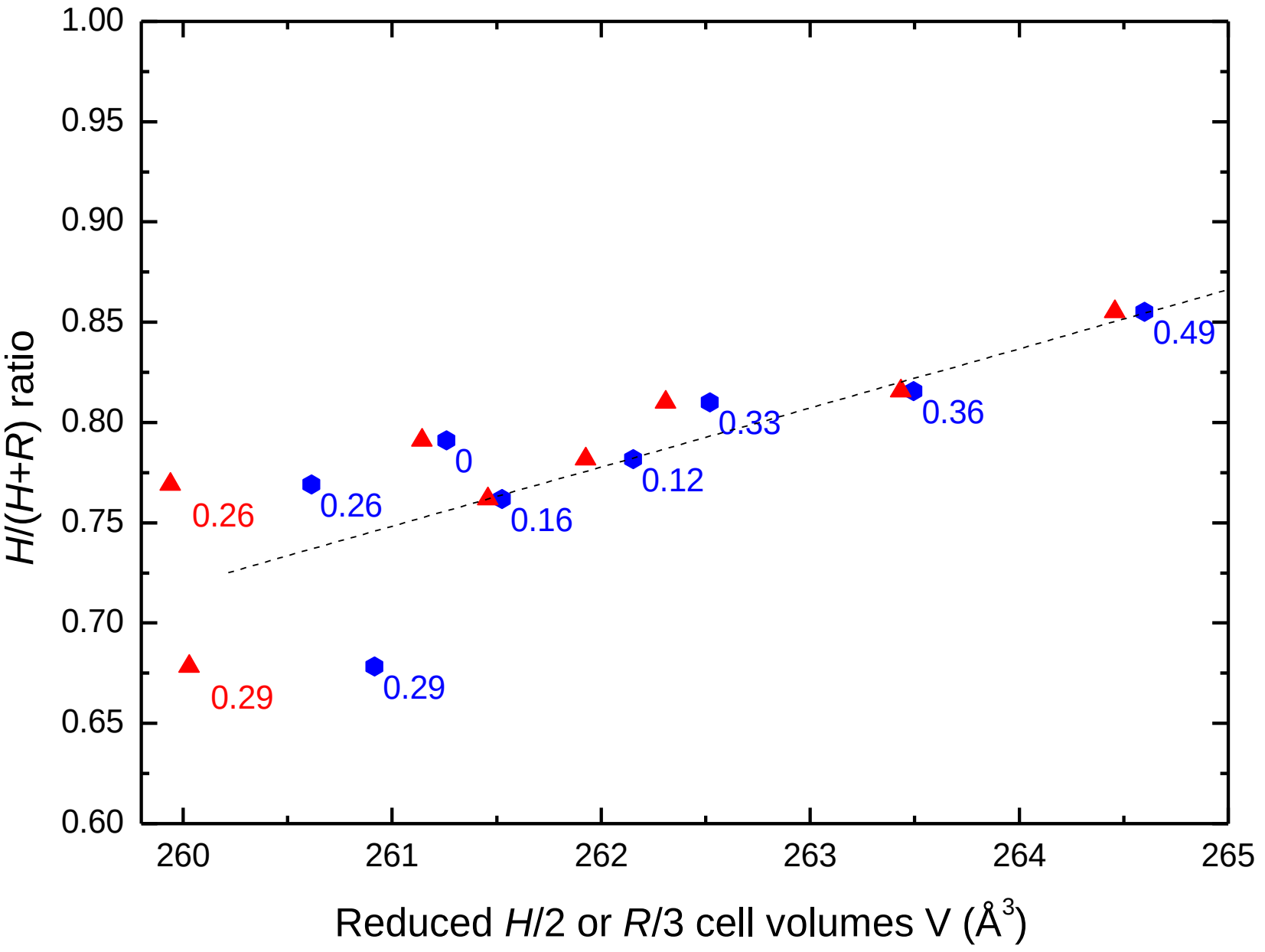

**Figure 2.** Evolution of the $H/(H + R)$ ratio (%) as a function of the reduced cell volumes ($V_H/2$ (blue circle) and $V_R/3$ (red triangle) for $Sm_2Mn_xNi_{7-x}$ ($0 \leq x \leq 0.49$). The x values are given directly on the graph. The dotted line represents a linear adjustment of the data.

Figure 3 shows the evolution of the cell volumes and lattice parameters for the *R* and *H* phases. Despite a continuous increase of the Mn content, the cell volume and parameters do not follow a linear variation as can be expected from a typical Vegard's law [32]. Though the atomic radius of Mn is larger than that of Ni, *V*, *a* and *c* decrease between $x = 0.12$ and $x = 0.26$, and then increase between $x = 0.26$ and $x = 0.49$. The same behavior is observed for the *R* and *H* phases. Variation of the Sm stoichiometry as a function of *x*, obtained from EPMA and XRD analyses, is also shown for the sake of comparison.

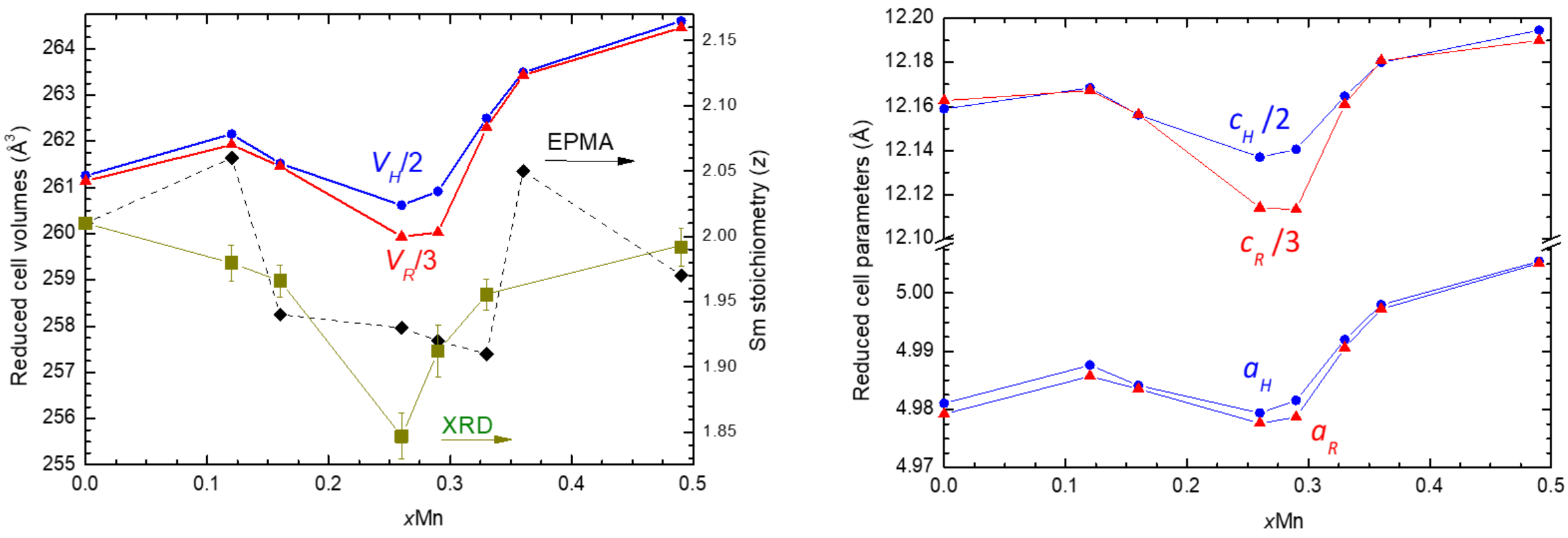

**Figure 3.** (**left frame**) Evolution of the reduced cell volumes ($V_H/2$ (blue circle) and $V_R/3$ (red triangle) (left axis) and Sm concentration obtained by EPMA (black diamond) and XRD analysis (green square)(right axis) as a function of *x* for $Sm_2Mn_xNi_{7-x}$; (**right frame**) Individual variations of the reduced cell parameters ($a_H$, $c_H/2$, $a_R$, $c_R/3$).

### 3.2. $H_2$ Sorption Properties

The *P-c* Isotherms of $Sm_2Mn_xNi_{7-x}$ with $x = 0$, 0.12, 0.29 and 0.33 were measured at 25 °C and are shown in Figure 4. At the first cycle, each compound exhibits at least two plateaus upon absorption but different behaviors for desorption. The absorption plateau pressures, capacities and desorption ratio measured during the first hydrogenation cycle are given in Table 2.

Compared to binary $Sm_2Ni_7$, the substitution of nickel by a tiny amount of manganese strongly modifies the *P-c* isotherm shapes. First, all Mn-containing samples form a larger solid solution with hydrogen up to 1 H/f.u. (α phase). Second, the plateau pressures change drastically. With $x_{Mn}$ growing, the first plateau pressure globally increases (except for $x = 0.29$ for which the secondary $AB_5$ phase (6 wt %; $P_{eq} \approx 0.1$ MPa) might shift the plateau pressure), whereas the second plateau pressure strongly decreases. Accordingly, the gap between the two plateaus is strongly reduced leading to higher reversible hydrogen concentration in a narrower pressure window. As concerns the overall capacity (given at 10 MPa in Table 2, it rises with Mn content up to 1.47 wt % for $x = 0.29$ then slightly decreases to 1.41 wt % for $x = 0.33$. In addition, the practical capacity is greatly improved due to smaller hysteresis and much better reversibility of the hydrogenation reaction, reaching 96% for $x$ >0.3 from the first cycle.

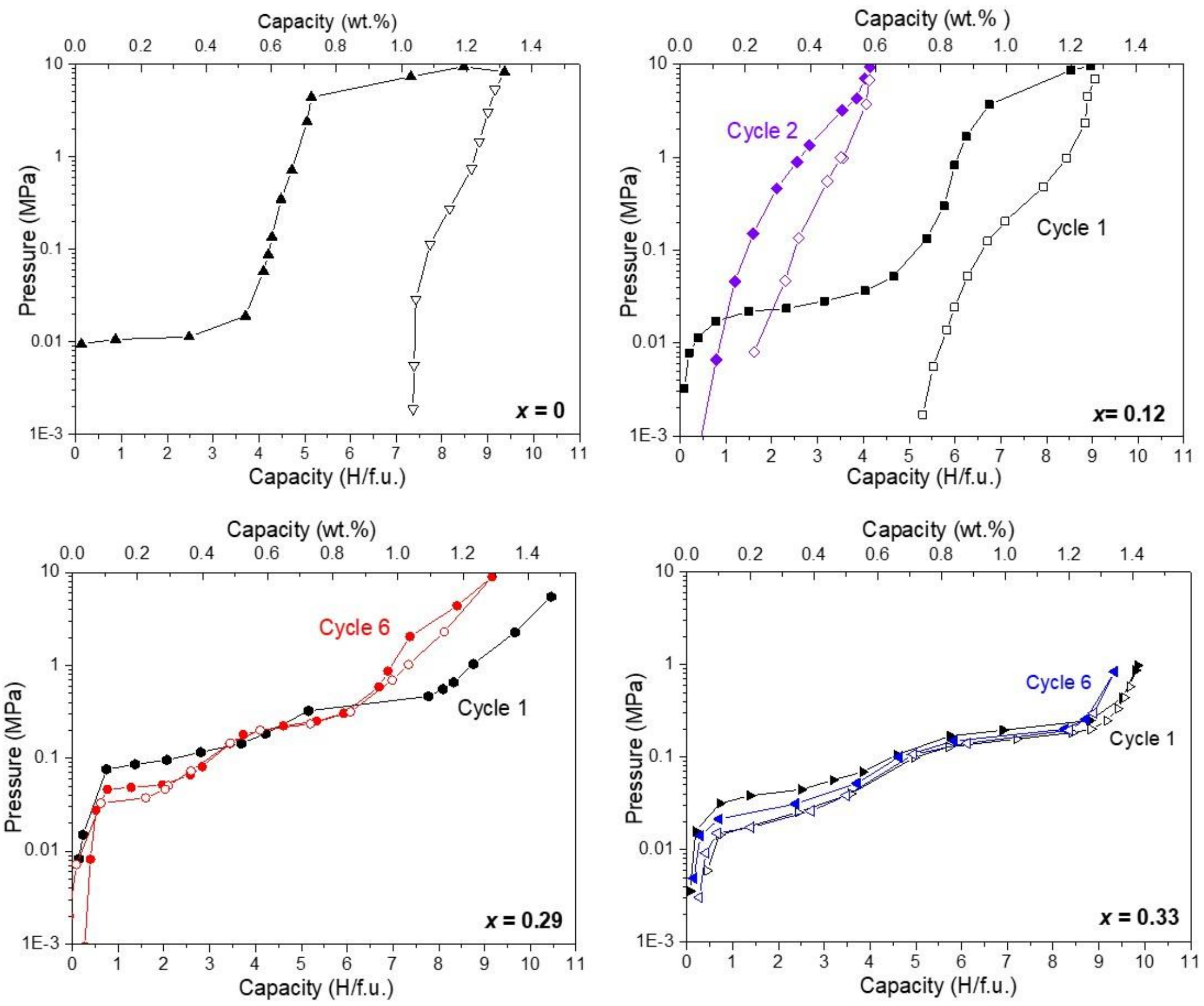

**Figure 4.** P-c isotherms of $Sm_2Mn_xNi_{7-x}$ measured at 25 °C with $x = 0$, 0.12, 0.29 and 0.33, full symbols stand for absorption, open symbols stand for desorption.

**Table 2.** First hydrogenation cycle properties at 25 °C of $Sm_2Mn_xNi_{7-x}$ compounds with x = 0, 0.12, 0.29 * and 0.33.

| $Sm_2Mn_xNi_{7-x}$ | $P_{eq}$ abs. (MPa) | | Total Capacity (10 MPa) | | Desorption Capacity (%) |
|---|---|---|---|---|---|
| | 1st $P_{eq}$ | 2nd $P_{eq}$ | (H/f.u.) | (wt.%) | |
| **0** | 0.011 | 6.5 | 8.5 | 1.19 | 24 |
| 0.12 | 0.027 | 6.0 | 9.0 | 1.25 | 41 |
| 0.29 * | 0.1 | 0.39 | 10.5 | 1.47 | 86 |
| 0.33 | 0.041 | 0.21 | 9.8 | 1.41 | 96 |

(* containing 6 wt % of $SmNi_5$-type phase, see Table 1).

For each hydrogenated sample ($x$ = 0, 0.12 0.29 and 0.33), XRD characterization was performed after hydrogenation cycling at 10 MPa to observe any structural changes occurring during the process. Because of their poor reversibility, XRD analysis for $x$ = 0 and 0.12 were performed after the first cycle. For $x$ = 0.29 and 0.33, showing better reversibility, XRD patterns were measured after 6 and 10 cycles, respectively.

The diffractograms obtained after dehydrogenation are compared to the pristine materials (Figure S2). For $x$ = 0 and 0.12, the structure drastically evolves from the very first hydrogenation cycle. The diffraction peaks are broader indicating a loss of crystallinity (decrease in crystallite size and/or increase in structural defects). For $x$ = 0.29 and 0.33, after 6 and 10 cycles respectively, the initial structure is fully retained without noticeable modification, attesting for the good reversibility of these materials upon cycling.

### *3.3. Calendar Corrosion*

To study the influence of the manganese on the corrosion process, samples with $x$ = 0, 0.12 and 0.26 were immersed in a solution of KOH 8.7 M for different durations ($t$ = 24 h, 1 w, 4 w, 8 w and 18 w). After rinsing and drying, the corroded samples were analyzed by XRD. The patterns are presented in Figure S3. The strong diffraction peak at 16.05° is characteristic of the $Sm(OH)_3$ hydroxide. The relative intensities of the hydroxide peaks increase with the corrosion time for all materials indicating a corrosion evolution. Corrosion rates depend of the Mn content. Indeed, $Sm(OH)_3$ appears after 24 h for the binary compound, only after 1 w for the substituted ones, and the binary alloy corrodes faster than the substituted alloys. Formation of $Sm(OH)_3$ is lower for $x$ = 0.12 than for $x$ = 0 and $x$ = 0.26, for corrosion times over 4 w. Apart from $Sm(OH)_3$, formation of metallic nickel is also observed for long corrosion times.

Whatever the value for $x$ (0, 0.12, 0.26), needle-like corrosion products are clearly observed by SEM after 24 h (Figure 5 and S4). They start to grow on preferred nucleation areas circled in green in Figure 5.a, b, where numerous small petal-shaped seeds (about 150 nm in diameter and 20 nm thick) are concentrated. The 2-µm long and 400-nm large needles are observed in these areas and many other places, and their size and number increase with time. These needles have been already reported in the literature [29–31,33,34] and are attributed to rare earth hydroxides. However, Mn modifies the corrosion mechanism as two other morphologies are visible for Mn-containing samples. First the area framed in red in Figure 5.a shows no germs or needles for $Sm(OH)_3$. Second, the area circled in blue (Figure 5.a, c) show porosities of ~200 nm, only observed in Mn-containing compounds. For long-time corroded samples (18 w for $x$ = 0, 0.12; 8 w and 18 w for $x$ = 0.26, Figure S5), a few corrosion products with hexagonal shapes can be observed locally in some SEM micrographs.

EDS elemental maps have been recorded on a cross section of a grain corroded 8 weeks for $x$ = 0.26 (Figure S6). They highlight the spatial distribution of the corrosion products, with a nickel layer enriched in oxygen right at the alloy surface, and samarium hydroxide needles on top of this nickel layer. The nickel layer enriched in oxygen corresponds to metallic nickel nanocrystallites surrounded by a nanometric layer of NiO, as it has been observed using the Bias effect measurement in previous studies on $Y_2Ni_7$ alloys [27]. In the present work, the strong magnetocrystalline anisotropy of Sm impedes such Bias effect evaluation.

In addition, large, flat and porous corrosion products are also found apart from the grains for Mn-containing samples (Figure 5.d, e). EDS measurements done on these corrosion products give an average composition $K_{11}Mn_{25}O_{64}$. This corrosion product, though noteworthy, does not appear in the X-ray diffractograms and seems therefore poorly crystallized. For a better identification of this phase, Raman micro-spectroscopy was used (Figure 5.f) and comparison of the measured spectrum with literature data leads to the identification of a birnessite-type oxide $K_{0.5}Mn_2O_4.nH_2O$ [35].

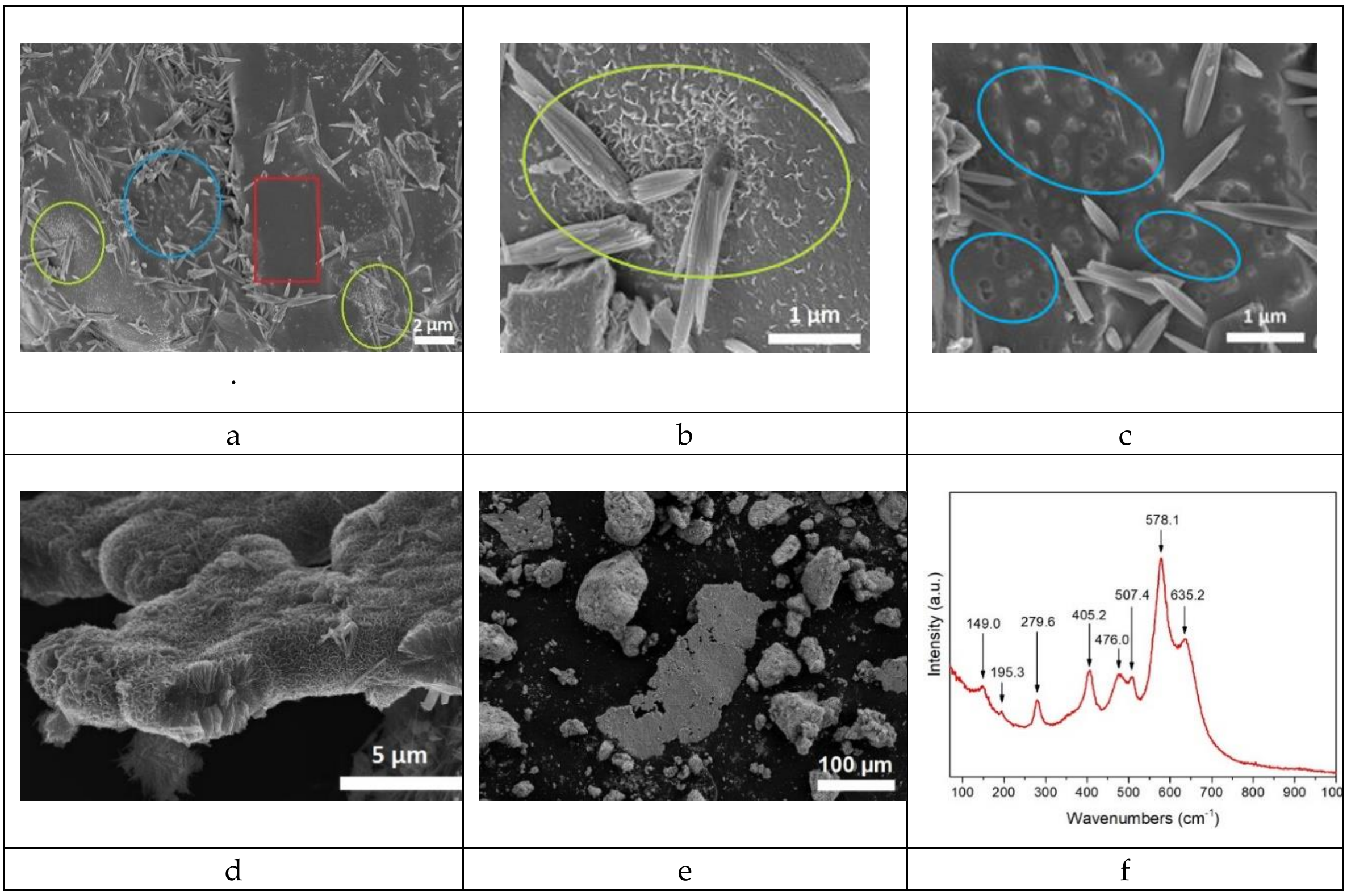

**Figure 5.** SEM micrographs of (**a**–**c**) the surface of 24 h-corroded grains: (**a**) $x = 0.12$; (**b**,**c**) $x = 0.26$. (**d**,**e**) view of a corrosion product aside from a grain for (**d**) $x = 0.26$ corroded 4 weeks and (**e**) 18 weeks. Three different morphologies are visible: Blue circles: porosities; Green circles: preferred nucleation with petal-shaped seeds; Red area: no germs or needles. (**f**) Raman spectra of the Mn-rich phase (after 18 weeks) identified as K-birnessite (frequencies reported by arrows are attributed to birnessite by comparison with [35]).

Regarding the kinetic point of view, two different methods have been used to assess the corrosion progress. The first one is based on the measurements of the magnetization of the corroded samples ($x = 0$, 0.12 and 0.26) as a function of the magnetic field (Figure S7). This method assumes that all other species (parent alloy, Sm hydroxide, Mn oxide) are paramagnetic at room temperature and contribute only to the slope of the $M(H)$ curves. It is also assumed that the Ni particles are nanoporous, large enough to display a ferromagnetic behavior and not modified by a possible Mn solid solution as the relative Mn content is very small. Free metallic np-Ni has been then quantified from the saturation magnetization at 300 K and the results are shown in Figure 6.

Whatever the corrosion time, the Mn-free binary compound contains the larger free nickel amount with values above 11 wt % after 8 weeks. As concerns the Mn substituted alloys, the amount of free nickel is lower (below 10 wt %) and varies with the composition and corrosion duration $t$. At short corrosion time ($0 < t < 4$ weeks), the Ni formation is comparable for both materials ($x = 0.12$ and 0.26). At the intermediate time ($4 < t < 8$ weeks), the Mn-rich alloy ($x = 0.26$) corrodes faster than $x = 0.12$. For the longer time ($t \geq 8$ weeks), surprisingly, the amount of Ni grows slowly for $x = 0$ and 0.12 but decreases slightly for $x = 0.26$. Likely, for this concentration, part of the formed nickel transformed in other species, no longer participating in the saturation magnetization.

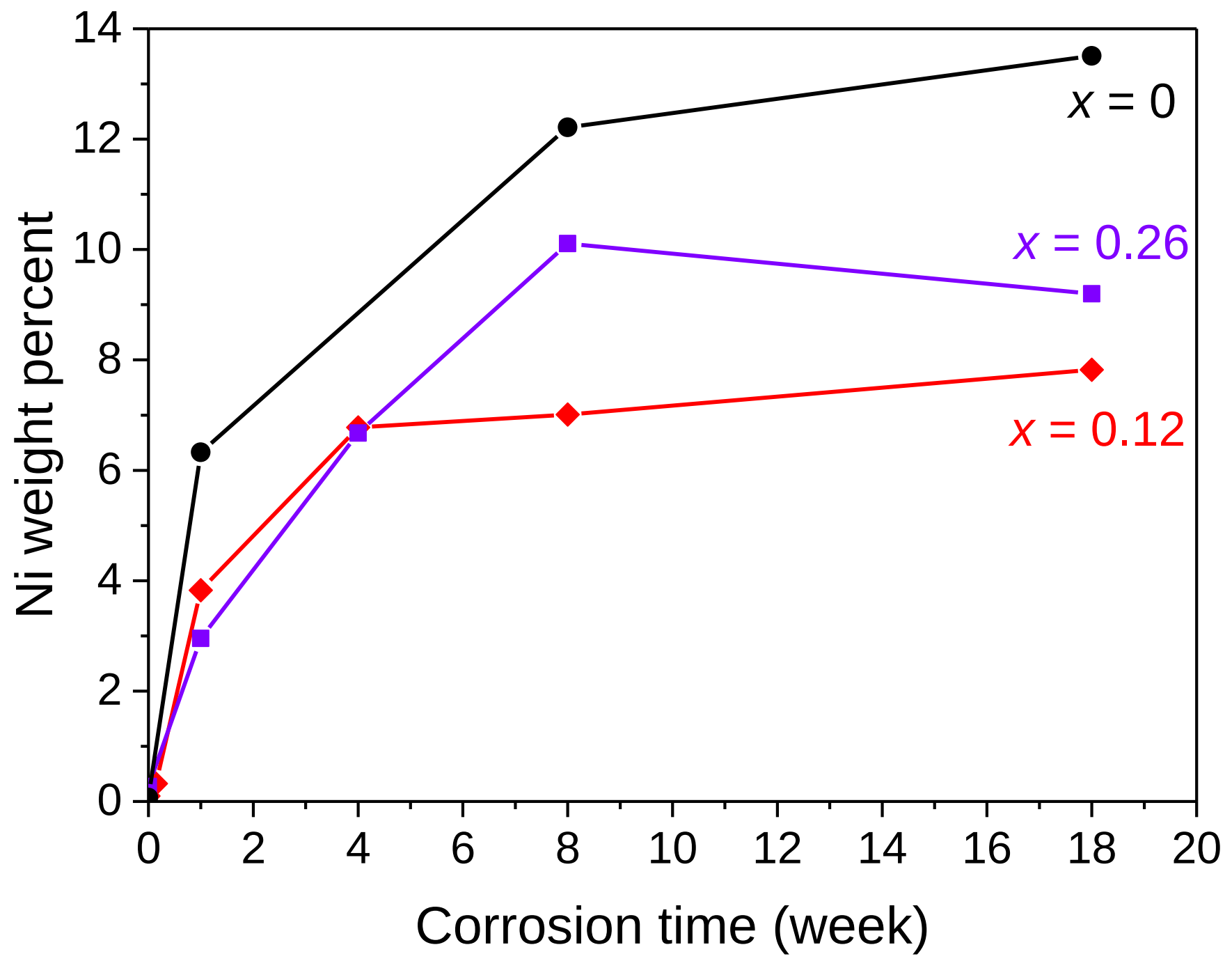

**Figure 6.** Weight percent of nickel formed as a function of corrosion time for $Sm_2Mn_xNi_{7-x}$ ($x$ = 0: black; 0.12: red; 0.26: purple).

TGA was the second method used to access to the corrosion rate. The results of the TGA analysis for $x$ = 0, 0.12 and 0.26 corroded between 8 and 18 weeks are shown in Figure 7.

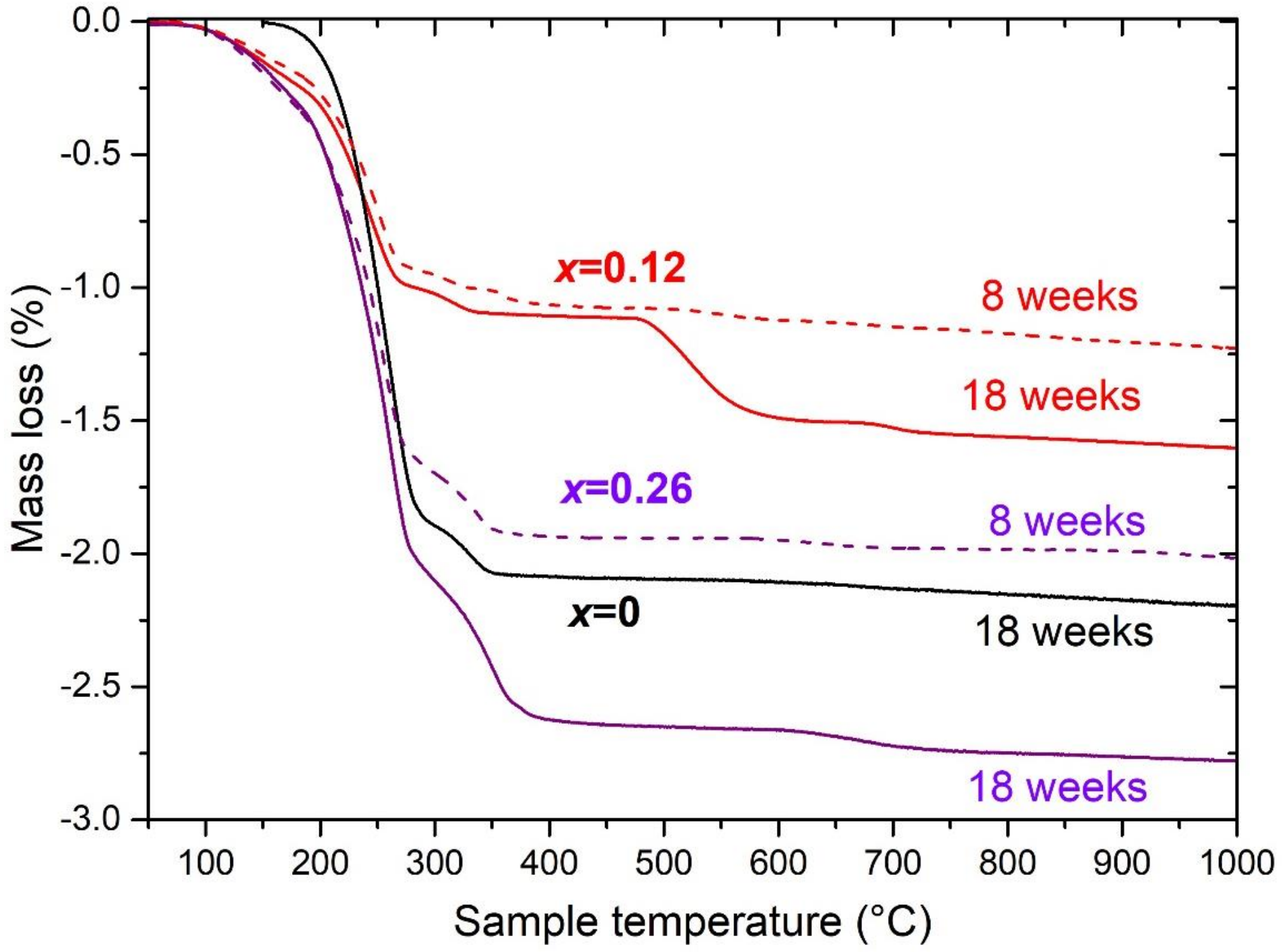

**Figure 7.** TGA of $Sm_2Mn_xNi_{7-x}$ for $x$ = 0 (black), $x$ = 0.12 (red) and $x$ = 0.26 (purple); after 8 weeks (dotted line) and 18 weeks (full line) of corrosion.

The mass loss curves show several steps. Assuming that the dehydration of $Sm(OH)_3$ is mainly responsible for the mass losses observed between 150 °C and 450 °C, one can estimate according to Equations (1) and (2), the amount of formed hydroxide. For temperatures higher than 450 °C, weight losses can be attributed to oxygen loss of the birnessite [36,37] and/or to the decarbonation of some carbonated-$Sm(OH)_3$ formed during long-term storage of the corroded alloys [28]. The quantitative mass losses and the wt % of Ni obtained from the TGA curves applying Equation (3), are given in Table 3.

**Table 3.** TGA mass loss and estimated wt % of nickel obtained from this mass loss if attributed to $Sm(OH)_3$ amount, compared to the wt.% of nickel measured from magnetic measurements. The difference between the Ni wt % given by the two techniques enables us to estimate the $Ni(OH)_2$ wt % in the corroded samples, and deduce the weight percent of $Sm(OH)_3$ in the corroded samples.

| $Sm_2Mn_xNi_{7-x}$ | Corrosion Time (weeks) | TGA Mass Loss [150–450 °C] (±0.3 wt %) | Ni Calculated from TGA Loss (±0.3 wt %) | Metallic Ni Measured from Magnetism (±0.2 wt %) | $Ni(OH)_2$ from the Difference between TGA and Magnetism (±0.5 wt %) | $Sm(OH)_3$ from the Difference between TGA and Magnetism (±0.5 wt %) |
|---|---|---|---|---|---|---|
| 0 | 18 | 15.6 | 15.9 | 13.5 | 0.8 | 13.8 |
| 0.12 | 8 | 6.4 | 6.2 | 7 | 0 | 7 |
| 0.12 | 18 | 8.4 | 8.2 | 7.8 | 0.1 | 8.0 |
| 0.26 | 8 | 12.2 | 12.4 | 10.1 | 1.2 | 9.6 |
| 0.26 | 18 | 17.6 | 17.9 | 9.2 | 3.3 | 10.7 |

## 4. Discussion

### *4.1. Evolution of the Structural Properties*

All compounds form the *R* and *H* polymorphs as a main $A_2B_7$ phase in the studied range of substitution ($0 \le x < 0.5$). Beside the formation of low quantity of oxides, small amounts (<6.5 wt %) of the neighboring phase $SmNi_5$ are observed for $x = 0.26$ and 0.29. More significantly, the $SmNi_3$-type phase appears in larger quantities (>24 wt %) for $x \ge 0.36$. These observations are in line with the ternary diagram at 400/600 °C published by Bodak [38] showing that above 4 at.% Mn, the 2:7 phase coexists with the $AB_3$- and $AB_5$-type neighboring phases. In the present work, *x* values remain just below the 4 at.% Mn limit but the synthesis was made at 800/900 °C.

As commonly observed for the 2:7 stoichiometry, the two polymorphs (*R* and *H*) coexist for all *x* values. The occurrence of these two polymorphs is generally related to thermal history and cell volume effects [25]. It has been reported that a smaller radius of the rare-earth favors the formation of the *R* structure. Thus, substitution of Ni for Mn should promote the *H* structure as Mn being larger than Ni, will induce cell volume expansion. This is indeed the case looking at the $H/(H + R)$ ratio as a function of the reduced cell volumes (i.e., reduced to $V_H/2$ and $V_R/3$ for sake of comparison) shown in Figure 2. Beside small deviations, the studied phases show a linear increase of the $H/(H + R)$ ratio with reduced cell volume, confirming the geometric effect.

It is also worth noting that the *H* and *R* cell volumes vary the same way (Figure 3), indicating that Mn substitution is nearly equally distributed in both polymorphs. However, analysis of the structural data shows that the cell volume expansion does not follow a typical linear behavior. Indeed, the cell volume increases for low values ($0 \le x \le 0.11$), then starts to decrease ($0.11 < x \le 0.26$) to reach a minimum value lower than for $Sm_2Ni_7$ itself (i.e., $x = 0$). Larger $x$ values lead to a sharp volume increase ($0.26 < x \leq 0.36$) to finally recover a progression ($0.33 < x \leq 0.49$) comparable to that of lower $x$ values ($0 \leq x \leq 0.12$). Anisotropic cell parameter variation has already been reported in these stacking structures and attributed to uneven distribution of the substituting element between the two sub-units ([$AB_2$] and [$AB_5$]) [29,39] leading to dissimilar variations along the c-axis with respect to the basal *a*-*b* plane.

However, in the present case, the variations are rather uncommon as the substitution induces successive elongation and contraction of the cell, leading to cell volumes even smaller than the starting binary phase. In addition, volume evolution remains isotropic as the variations of *a* and *c* are simultaneous. This intriguing behavior can be understood considering two hypotheses. First, some rearrangements of the Mn atom distribution within the different Ni sites might induce anomalous volume evolution, though a contraction beneath the volume of $Sm_2Ni_7$ can hardly be realized. The second hypothesis is related to the possible formation of defects such as rare earth vacancy to accommodate the volume increase induced by Mn substitution. This hypothesis is supported by the EPMA and XRD measurements showing a likely correlation between the Sm concentration and the cell volume evolution (Figure 3). Rare earth vacancy occurrence is known to appear in some Laves phases [40–42], corresponding here to the [$AB_2$] sub-units. Similarly, a large domain of existence has been reported for $LaNi_5$ (up to $LaNi_{5.4}$) [43,44], corresponding here to the $AB_5$ sub-units. This large domain was accounted by the substitution of Ni dumbbells at the La location leading to the stoichiometry $La_{1-\eta}Ni_{5+2\eta}$ and inducing a cell volume contraction with $\eta$. To our knowledge, such defects (either vacancy formation or dumbbells substitution) have never been reported for $A_2B_7$-type compounds but cannot be excluded in the present case. Finally, it is worth noting that for the highest $x$ values ($x \geq 0.33$), the system seems recovering a linear behavior close to a Vegard's law, indicating that the defects have been mitigated in the structure and that the geometric effect is the main one for these compositions. Unfortunately, the discrimination of Mn toward Ni cannot be done by standard XRD, due to the poor contrast between the two atoms, nor by neutron diffraction according to the very high absorption of Sm. Only high resolution anomalous diffraction at the K edges of Mn and/or Ni might shed light on the repartition of these 3*d* metals within the available sites and could confirm the formation of vacancies or dumbbells in these stacking structures.

### 4.2. Hydrogen Sorption Properties

Looking to the *P*-*c* isotherm curves, the Mn substitution brings decisive changes in the hydrogen sorption properties. First, whereas the system is poorly reversible for $0 \leq x \leq 0.12$ (Figure 4), an improved behavior is obtained for larger $x$ values. This is attributed to the better stability of the crystal structure upon hydrogenation as attested by the preserved crystallinity of the metallic phase after dehydrogenation for higher $x$ values (Figure S2).

Regarding capacity, adding a small quantity of Mn slightly increases the absorbed hydrogen capacity and considerably improves the desorption ability, as well as the cyclability. The main improvement comes from the good reversibility and the reduced gap between the two plateaus, making the useable capacity much larger in a narrower pressure range. Indeed, the plateau pressure of the second plateau decreases notably with increasing amount of manganese, from 6.5 MPa for $x = 0$ down to 0.21 MPa for $x = 0.33$ (Figure 4). On the other hand, the first plateau slightly increases with $x_{Mn}$ (Table 2). Previous studies from Berezovets *et al.* [39,45] for the system $YMn_xNi_{3-x}$ indicates that Mn preferentially occupies the [$AB_5$] sub-unit leading to strongly improved reversibility. Similarly, a work by Yasuoka *et al.* on the substitution of nickel by aluminum in $Nd_{1.8}Mg_{0.2}Ni_7$ shows that this yields an increase volume of the [$AB_5$] sub-unit, thus reducing the size difference between the [$AB_5$] and [$AB_2$] sub-units, greatly improving the reversibility [15]. The atomic radius of manganese, though smaller than the Al one, is still larger than nickel and a similar mechanism is expected. This geometrical effect is further confirmed by several authors for various systems [20,29,46,47] and can be considered as a general rule allowing for optimizing the sorption properties of this family of hydrides. In the present case, $x \geq 0.29$ is required to induce this beneficial volume effect. However, larger values ($x > 0.33$) leads to the formation of the neighboring $AB_3$ phase detrimental for hydrogen reversibility so that an optimum is found in the range $0.29 \leq x \leq 0.33$.

### 4.3. Calendar Corrosion Mechanisms

Calendar corrosion in an alkaline medium of the $Sm_2Mn_xNi_{7-x}$ pseudo-binary compounds ($x$ = 0.12 and 0.26) was studied and can be compared to $Sm_2Ni_7$ ($x$ = 0). The Mn-containing compounds show porosities of about 200 nm at the grain surface, observable after 24 h of corrosion (Figure 5.a, c). This can be attributed to a rapid dissolution of manganese as the solubility of Mn increases with pH [48]. Indeed, the manganese turbidity diagram [49] indicates that $Mn(OH)_3^-$ is the most stable ionic species under our calendar corrosion pH conditions. The dissolution of manganese to form $Mn(OH)_3^-$ will therefore lower the pH locally and the rapid dissolution of manganese seems to slow down the formation of rare earth hydroxide and metallic nickel. Indeed, magnetic measurements for $x$ = 0.12 and $x$ = 0.26 show a lower nickel percentage (3.8 and 3.0 wt % respectively, Figure 6) than for $x$ = 0 (6.3 wt %, Figure 6) after 1 week corrosion, so it is likely that the dissolution of manganese first slows down the formation of metallic nickel, and that of rare earth hydroxide. This effect is seen as a kind of sacrificial anode mechanism setting up at the early corrosion stage.

Dissolution of manganese lowers the pH and changes the morphology of the $Sm(OH)_3$ needles. As the pH is lowered, germination and crystallization of $Sm(OH)_3$ becomes more difficult, so crystallization of the hydroxides along the basal plane is favored. Following this idea, the larger the Mn amount in $Sm_2Mn_xNi_{7-x}$, the longer and finer the needles of $Sm(OH)_3$. Indeed, after 1 week of corrosion (Figure S4), differences in morphology are clearly observed. $Sm(OH)_3$ needles are thicker for $x$ = 0.12 (about 1 μm) than for $x$ = 0.26, (only 100 nm wide). On the other hand, the density of needles at the surface is larger for $x$ = 0.26 than for $x$ = 0.12.

The presence of manganese also induces the presence of a new corrosion product. Large porous platelets containing manganese, oxygen and potassium are observed for all pseudo-binaries. These platelets, rich in potassium, become easily visible after 4 weeks of corrosion (Figure 5.d, e). They form in solution and not directly on the grain surface. Raman micro-spectroscopy allows for identifying these platelets as K-birnessite, a lamellar manganese oxide of formula $K_xMnO_2.nH_2O$.

From the TGA data, and assuming that the weight loss in the [150–450]°C temperature range are due to $Sm(OH)_3$ dehydration, the weight percent of the corrosion products $Sm(OH)_3$ and metallic Ni have been calculated and are summarized and compared to the magnetic measurements in the third and fourth columns of Table 3. Based on TGAs and magnetic measurements, corrosion rate is slowed down relative to the binary alloy for low Mn amount ($x$ = 0.12). For this sample corroded 8 w, TGA yields nickel amount in very good agreement with magnetic measurements. However, for longer corrosion time or higher Mn amount ($x$ = 0.26), a significant difference between TGA and magnetic values is observed (Table 3). In addition, the quantity of metallic nickel decreases unexpectedly between 8 and 18 weeks for $x$ = 0.26 (Figure 6). This difference between Ni measured by magnetism and Ni estimated from TGA can be understood by considering that metallic nickel surface oxidizes into NiO, which then hydrolyses into $Ni(OH)_2$ as for the $La_2Ni_7$ system [34]. This hypothesis is in good agreement with the few corrosion products with an hexagonal shape observed for $x$ = 0.26 corroded 8 and 18 w, and for $x$ = 0 and 0.12 corroded 18 w (Figure S6), as $Ni(OH)_2$ generally crystalizes under this shape in the considered systems [34]. Moreover, $Ni(OH)_2$ will dehydrate during heating in TGA in the same temperature range than $Sm(OH)_3$, so will contribute to the steps initially attributed to $Sm(OH)_3$ only [50]. Assuming a negligible amount of NiO (as the oxidized layer is nanometric) and combining the Ni wt % from TGA with the Ni wt % from magnetism, the quantity of $Ni(OH)_2$ wt % can be estimated and the corrected quantity of $Sm(OH)_3$ can be deduced (Table 3). As a conclusion, considering this corrected quantity of hydroxide, although manganese slows down the corrosion rate compared to the binary alloy, for corrosion durations longer than four weeks $Sm_2Mn_{0.26}Ni_{6.74}$ corrodes faster than $Sm_2Mn_{0.12}Ni_{6.88}$. This highlights the complexity of the corrosion mechanisms: a sacrificial anode mechanism takes place with Mn, whereas Mn corrodes simultaneously to form the birnessite. Thus, Mn favors the hydrolysis of NiO into $Ni(OH)_2$, which increases the corrosion rate. It is therefore important to tune the Mn content carefully in these substituted compounds. A high Mn content enhances the corrosion, but a tiny amount will drastically improve the hydrogen sorption properties.

## 5. Conclusions

The pseudo-binary system $Sm_2Mn_xNi_{7-x}$ has been investigated for various $x$ values ($0 \leq x < 0.5$). For all compositions, it forms $A_2B_7$-type polymorphs (*R* or *H*) as a main phase. The Mn for Ni substitution does not lead to a linear increase of the cell parameters and volume but to unexpected variations than can be interpreted on the basis of structural defects (rare earth vacancies) and uneven distribution of the Mn atoms in the Ni sublattice and/or in the [$AB_5$], [$AB_2$] sublattices.

All compounds absorb a significant hydrogen quantity up to 10 MPa (>1.2 wt %). However, reversible capacity is strongly improved for $x$ values larger than 0.3, due to a better structural stability of the phases and moreover, to a strong reduction of the hysteresis and gaps between the two plateaus. This leads to larger and flatter isotherm curves allowing reversible storage of hydrogen >1.4 wt % at practical conditions (room temperature and atmospheric pressure). Such improvement is attributed to geometrical effects induced by the Mn substitution and reducing the size difference between the [$AB_5$] and [$AB_2$] sub-units, leading to enhance reversibility.

Beside improved capacity, resistance to corrosion is also a key parameter for these materials useable either for dihydrogen storage (that might contain water if produced by alkaline electrolysis) or for electrochemical storage in alkaline batteries. Calendar corrosion is strongly influenced by the presence of Mn in this pseudo-binary system. Dissolution of Mn is observed, inducing local pH depletion responsible for changing the morphology of the Sm hydroxide needles, slowing down the corrosion process and forming a new corrosion product identified as K-birnessite. Thus, Mn addition triggers a sacrificial anode mechanism taking place at the early corrosion stage.

**Supplementary Materials:** The following are available online at www.mdpi.com/xxx/s1, Figure S1: Full repartition of the EPMA measurements in the Mn-Ni-Sm ternary phase diagram for $Sm_2Mn_xNi_{7-x}$ ($0 \leq x \leq 0.49$). Iso-composition lines for $SmB_5$, $Sm_2B_7$ and $SmB_3$ are shown in red in wine). Figure S2: Diffraction patterns of $Sm_2Mn_xNi_{7-x}$ ($x$ = 0; 0,12; 0,29 et 0,33) for the pristine alloys (black lower curve) and after dehydrogenation under dynamic primary vacuum at 150 °C (upper red curve). Number of cycles are shown directly on the graph. Figure S3: Diffraction patterns for $x$ = 0 (left), $x$ = 0.12 (center) and $x$ = 0.26 (right) after different corrosion times in KOH 8.7 mol.$L^{-1}$. Formation of $Sm(OH)_3$ (black circle), and metallic nickel (red diamond) is observed upon time. Figure S4: SEM micrographs $Sm_2Mn_xNi_{7-x}$ after 1 week of corrosion; $x$ = 0, 0.12 and 0.26 (left to right). Figure S5: Left, SEM micrograph of $Sm_2Mn_{0.12}Ni_{6.88}$ after 18 weeks of corrosion (Inlense detector), and right, $Sm_2Mn_{0.26}Ni_{6.74}$ after 8 weeks of corrosion (SE2 detector) highlighting the local presence of hexagonal shaped corrosion products (circled in orange) attributed to $Ni(OH)_2$. Figure S6: SEM-BSE micrograph and EDS elemental map for O (red), Sm (purple) and Ni (cyan) of a grain cross-section of sample for $x$ = 0.26 corroded 8 weeks. Figure S7: Evolution of the magnetization curves (300 K) for $x$ = 0.12 and 0.26 as a function of the field for different corrosion times ($0 \leq t \leq 18$ weeks).The curves for $x$ = 0 have been published in [30]

**Author Contributions**

Conceptualization, Judith Monnier, Junxian Zhang and Michel Latroche; Formal analysis, Nicolas Madern, Véronique Charbonnier, Judith Monnier, Junxian Zhang, Valérie Paul-Boncour and Michel Latroche; Funding acquisition, Michel Latroche; Investigation, Nicolas Madern, Véronique Charbonnier, Judith Monnier, Junxian Zhang, Valérie Paul-Boncour and Michel Latroche; Methodology, Nicolas Madern, Véronique Charbonnier, Judith Monnier, Junxian Zhang, Valérie Paul-Boncour and Michel Latroche; Project administration, Michel Latroche; Supervision, Judith Monnier, Junxian Zhang and Michel Latroche; Writing – original draft, Nicolas Madern, Judith Monnier, Junxian Zhang, Valérie Paul-Boncour and Michel Latroche; Writing – review & editing, Véronique Charbonnier.

**Funding:** This research received no external funding

**Acknowledgments:** The authors are thankful to E. Leroy for technical assistance in EPMA analysis. The authors are grateful to R. Baddour-Hadjean for her help to analyze the Raman spectra of K-birnessite.

**Conflicts of Interest:** The authors declare no conflicts of interest

Graphical abstract :

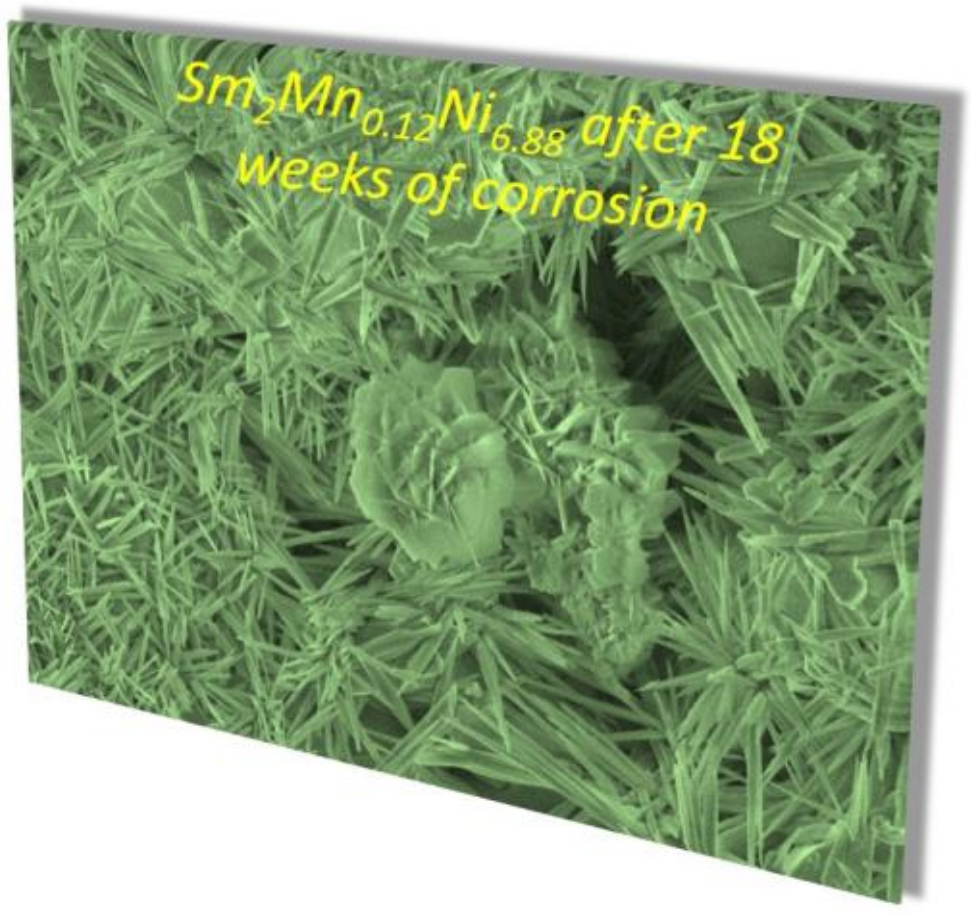